\documentclass[manuscript,screen,dvipsnames]{acmart}
\usepackage{graphicx}
\usepackage{subfigure}
\usepackage{makecell}
\usepackage{algorithm}
\usepackage{algorithmicx}
\usepackage[noend]{algpseudocode}
\usepackage{soul}
\usepackage{multirow}
\usepackage{rotating}
\usepackage{bigstrut}
\usepackage{url}
\usepackage{balance}
\usepackage{comment}

\newcommand{\tabincell}[2]{\begin{tabular}{@{}#1@{}}#2\end{tabular}}
\newcommand{\tool}{iLead}

\AtBeginDocument{%
  \providecommand\BibTeX{{%
    \normalfont B\kern-0.5em{\scshape i\kern-0.25em b}\kern-0.8em\TeX}}}

\setcopyright{acmcopyright}
\copyrightyear{2021}
\acmYear{2021}
\acmDOI{10.1145/xxxxxxx.xxxxxxx}

\acmJournal{JACM}
\acmVolume{37}
\acmNumber{4}
\acmArticle{111}
\acmMonth{8}

\begin{document}

\title{Identifying Emergent Leadership in OSS Projects Based on Communication Styles}


\author{Yuekai Huang}
\email{huangyuekai18@mails.ucas.ac.cn}
\affiliation{%
  \institution{Laboratory for Internet Software Technologies, Institute of Software Chinese Academy of Sciences; University of Chinese Academy of Sciences, Beijing}
  \country{China}
}

\author{Ye Yang}
\authornote{Corresponding author}
\email{yyang4@stevens.edu}
\affiliation{%
  \institution{School of Systems and Enterprises, Stevens Institute of Technology, Hoboken, NJ}
  \country{USA}
}

\author{Junjie Wang}
\email{junjie@iscas.ac.cn}
\authornotemark[1]
\affiliation{%
  \institution{Laboratory for Internet Software Technologies, State Key Laboratory of Computer Sciences, Institute of Software Chinese Academy of Sciences; University of Chinese Academy of Sciences, Beijing}
  \country{China}
}

\author{Wei Zheng}
\email{wzheng11@stevens.edu}
\affiliation{%
  \institution{School of Business, Stevens Institute of Technology, Hoboken, NJ}
  \country{USA}
}

\author{Qing Wang}
\email{wq@iscas.ac.cn}
\authornotemark[1]
\affiliation{%
  \institution{Laboratory for Internet Software Technologies, State Key Laboratory of Computer Sciences, Institute of Software Chinese Academy of Sciences; University of Chinese Academy of Sciences, Beijing}
  \country{China}
}

\renewcommand{\shortauthors}{Huang and Yang, et al.}

\begin{abstract}
In open source software (OSS) communities, existing leadership indicators are dominantly measured by code contribution or community influence. Recent studies on emergent leadership shed light on additional dimensions such as intellectual stimulation in collaborative communications. To that end, this paper proposes an automated approach, named iLead, to mine communication styles and identify emergent leadership behaviors in OSS communities, using issue comments data. We start with the construction of 6 categories of leadership behaviors based on existing leadership studies. Then, we manually label leadership behaviors in 10,000 issue comments from 10 OSS projects, and extract 304 heuristic linguistic patterns which represent different types of emergent leadership behaviors in flexible and concise manners. Next, an automated algorithm is developed to merge and consolidate different pattern sets extracted from multiple projects into a final pattern ranking list, which can be applied for the automatic leadership identification. The evaluation results show that {\tool} can achieve a median precision of 0.82 and recall of 0.78, outperforming ten machine/deep learning baselines. To demonstrate practical usefulness, we also conduct empirical analysis and human evaluation of the identified leadership behaviors from {\tool}. We argue that emergent leadership behaviors in issue discussion should be taken into consideration to broaden existing OSS leadership viewpoints.  Practical insights on community building and leadership skill development are offered for OSS community and individual developers, respectively.  

\end{abstract}




\maketitle

\section{Introduction}
\label{sec:introduction}

Steady influx of developer resource is key to healthy and sustainable OSS communities \cite{DBLP:journals/ijecommerce/HarsO02,DBLP:journals/ets/BaytiyehP10}. The presence of community leaders, who both steer the development direction and motivate more developers to contribute is crucial to ensure a successful outcome of an OSS project \cite{Li2012leadership}. In addition, it is also reported that developer retention is highly associated with the attention and treatment they receive in the project \cite{DBLP:conf/icse/ZhouM12,DBLP:conf/hicss/JensenKK11,DBLP:conf/sbsc/SteinmacherWCG12}. Therefore, it is an essential function for an OSS community to effectively monitor and appropriately recognize developer contribution in timely and fair manners.

In light of the importance of having leadership figures, many studies have investigated various developer leadership factors across OSS communities. Most rely on metrics representing individuals' code contribution or community influence. Example metrics of code contribution include code size (e.g., \#lines of code added/removed \cite{mockus2000case}, \#commits \cite{ben2013mining, da2014unveiling}), code structure (e.g., code instability \cite{ben2013mining}), code quality (e.g. \#problems reported \cite{mockus2000case}, \#buggy commits \cite{da2014unveiling}), or code topics \cite{linstead2007mining}. Example metrics of community influence include the number of followers (e.g., \#followers \cite{moqri2018effect, Li2012leadership}) and influence on other developer's motivations \cite{roberts2006understanding}. 

Most of today's OSS communities operate beneath an umbrella organization, 
while others are organized entirely independently, and  yet  others  follow a strategy somewhere  in  between \cite{eckert2019alone}. Different OSS organizational forms correspond with different leadership structure and coordination strategies, which complicates the measurement of leadership effectiveness. As detailed in Section \ref{sec:motivation}, we find that there is mismatch between these developers' contribution in issue discussion and their contributions measured by other dimensions such as code contribution or community followers. We conjecture that in OSS communities where development is highly autonomous, and explicit coordination and control are loose or non-existent, those who know how to communicate effectively with others will contribute more in leading, coordinating, and influencing the development. This indicates the need of a new lens in identifying and recognizing OSS leadership behaviors corresponding to more effective and efficient issue resolution. 

To bridge the gap, we start with constructing six categories of emergent leadership behaviors applicable to OSS projects from existing leadership models \cite{Li2012leadership, bass1996multifactor, zhu2012effectiveness, carte2006emergent, yukl2002hierarchical, yukl2012effective}. These include proposal, redirection, confirmation, inquiry, operation, and volunteer (as show in Table \ref{tab:cmt_types}). And then we propose an automated approach, named {\tool}, to mine communication styles and identify such emergent leadership behaviors in OSS communities, using issue comments data. The construction of {\tool} consists of three steps. First, we carry out the data labelling process to identify the category of leadership per issue comment. Second, we adapt the definition of linguistic patterns in previous studies \cite{DBLP:conf/europlop/Silva17,DBLP:conf/icse/0008ZBPL19,DBLP:conf/kbse/ShiCWLB17} and manually extract the heuristic linguistic patterns to express different types of communication styles in flexible and concise manners. Third, based on the two previous steps, we develop an automatic algorithm to merge and consolidate different pattern sets extracted from multiple projects into a final pattern ranking list. When applying {\tool} for leadership identification, the final pattern ranking list can be used to automatically match new issue comments and identify corresponding leadership behaviors. The evaluation is conducted on 10 popular OSS projects with 10,000 comments. Results show that {\tool} can achieve a median precision of 0.82 and recall of 0.78, outperforming ten machine/deep learning baselines. 

To demonstrate practical usefulness, we also conduct  empirical analysis and human evaluation of the identified leadership behaviors with {\tool}.
Results motivate the need for considering the emergent leadership behaviors in issue discussion to broaden existing OSS leadership viewpoints. Practical insights on community building and leadership skill development for OSS community developers are also discussed. To summarize, the main contributions of this paper are as follows:

\begin{itemize}
\item The construction of 6 categories of emergent leadership behaviors in OSS projects from issue comments data, adapting existing leadership models into OSS context. This is the first taxonomy of the emergent leadership behaviors in issue discussions of OSS communities, which can motivate following-up studies in this direction.

\item The development of {\tool} (with 304 defined linguistic patterns and pattern consolidation algorithm) to automatically identify emergent leadership behaviors in OSS projects from issue comments data. 
To our knowledge, this is the first study on automatic mining of emergent leadership behaviors in OSS issue discussion communications.

\item The evaluation of {\tool} using 10,000 comments extracted from 10 popular OSS projects, achieving promising results; and empirical study for revisiting the leadership landscape in OSS projects.

\item {Publicly accessible dataset and source code\footnote{\url{https://github.com/20210827/iLead}} to facilitate the replication of our study and its application in other contexts.}

\end{itemize}

\section{Background and Motivation}
\label{sec:motivation}
\subsection{Leadership Models and Theories}
\label{subsec_background_leadership}
It is important to note the difference between ``leadership'' and ``leader'', in that leadership status is not necessarily based on an OSS community position or designated authority, i.e., leader status \cite{eckert2019alone, northouse2021leadership,vanderslice1988separating}. Instead, leaders emerge and earn their status through incremental influences and contributions to OSS communities. In the context of this study, we adopt the Yukl's definition of leadership as ``influence exerted over other people to guide, structure, and facilitate relationships in a group'' \cite{yukl1981leadership}. Existing work on leadership theories and styles can be largely categorized into two polarized yet important leadership styles: transformational and transnational, depending on followers’ behavioral response \cite{bass2003predicting}. Transformational leadership \cite{burns1978leadership} requires leaders to work with followers to stimulate the enthusiasm of subordinates by creating a vision and motivation. Transactional leadership \cite{hollander1978leadership} is a kind of leadership behavior pays more attention to organizational management and performance. However, both styles are more focusing on traditional organizations with clear management hierarchy and team structure, which are not directly applicable to OSS communities.

In recent years, a number of new theories characterizing leadership in virtual teams were proposed, such as shared leadership \cite{yukl1981leadership, carson2007shared} and emergent leadership \cite{yoo2004emergent}. 
Shared leadership \cite{pearce2002shared} is a leadership mechanism existing in a team with clear hierarchy, and it is a process in which individuals in the group influence each other and work together towards the goal. Emergent leadership \cite{yoo2004emergent} is a spontaneous leadership mechanism as an emergent phenomenon that develops over time through group processes. 

\subsection{OSS Community, Project and Developers}
An OSS community is an ecosystem comprising of a set of closely-related OSS projects, which draws expertise and contributions from  a pool of developers \cite{jergensen2011onion}. 
OSS developers usually assume certain roles by themselves according to their personal interest, rather than being assigned a task by someone else \cite{nakakoji2002evolution}. This freedom in assuming roles/responsibilities, while significantly differing from traditional organizations, necessarily nurtures the open innovation that drives the success of OSS. As examples of such open contribution, developers may invest effort on a specific OSS project instead of others, and he/she may choose to submit code contribution, or participate in issue discussions. As examples of developer impact measures, the number of commits and the number of followers typically reflect a developer's relative influence in the OSS community. Consequently, an OSS developer's role in an OSS project may vary significantly based on his/her interest, availability, and expertise. 
Nakakoji et. al. \cite{nakakoji2002evolution} proposed to classify OSS developers' roles into 8 types including project leader, core member, active developer, peripheral developer, bug fixer, bug reporter, reader, and passive user. The order implies the relative significance of their contribution. In addition, in measuring developer's contribution, they usually play greater weight on the developer's code contribution \cite{github2021contributor,DBLP:conf/esem/LeeC17}, i.e., the number of commits, which is readily accessible from OSS platforms such as GitHub or SourceForge.

\subsection{A Motivational Example}
\begin{figure}[!t]
  \begin{minipage}[t]{0.48\linewidth}
    \centering
    \includegraphics[height=4.3cm]{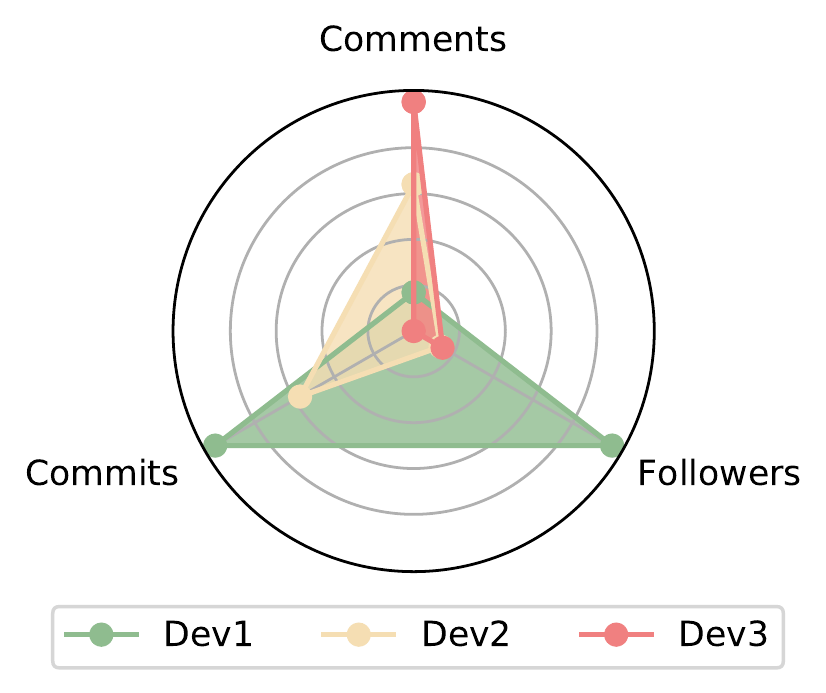}
    \caption{Three example developers in \emph{\textbf{atom}}}
    \label{fig:dev_examples}
  \end{minipage}
  \begin{minipage}[t]{0.48\linewidth}
    \centering
    \includegraphics[height=4.3cm]{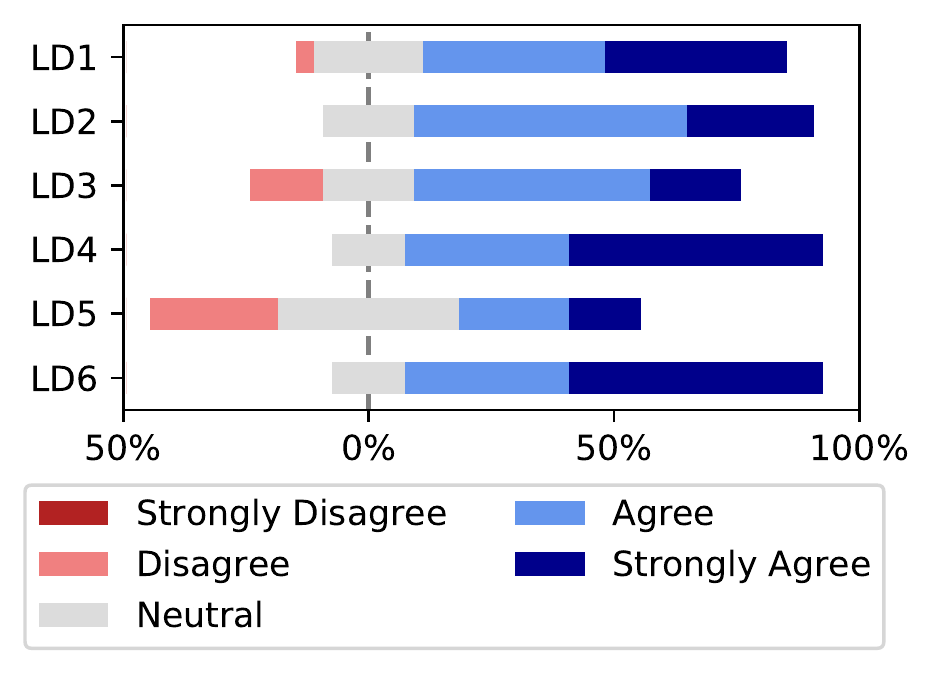}
    \caption{Survey results}
    \label{fig:survey}
  \end{minipage}
\end{figure}

To explore differences among different leadership indicators, 
we compare three indicators, i.e., \#commits, \#followers, and \#issue comments, across a set of contributing developers from the \textit{atom}\footnote{\url{https://github.com/atom/atom}} OSS project. Figure \ref{fig:dev_examples} illustrates the comparison results of three example developers. 
It is clear that developer \textit{Dev1} (id:1476) has the highest \#commits and \#followers among all three, followed by \textit{Dev2} (id:7910250) then \textit{Dev3} (ID:4525388). The first two developers, i.e., \emph{Dev1} and \emph{Dev2}, will likely be considered as having greater contributions than the third one following existing leadership indicators \cite{mockus2000case,ben2013mining,moqri2018effect, Li2012leadership} because they have more code contributions or have more followers. However, although developer \emph{Dev3} has less commits, he has actually published the most issue comments. Specifically, there are 267 comments made by \emph{Dev3}, including the following examples:

\begin{itemize}
\item \emph{``Have you tried safe mode? Try --safe and see if the issue stops.''}
\item \emph{``What is your atom version?''}
\item \emph{``Duplicate of \#1667''}
\end{itemize}

Through these communications, the developer \emph{Dev3} either provides alternative solution to the issue reporter, or asks clarification questions, or redirects the issue to other similar ones. These behaviors are actually representative of emergent leadership and managerial responsibilities,  which are essential to facilitate collaborative discussion and issue resolution. 

To further examine the differences of these indicators at project level, we rank all developers in \textit{atom} by the number of comments and the number of commits, respectively, and compare the top 25\% of the developers in the two rankings. Note that the number of followers is defined at the overall GitHub, therefore, we did not include it for the project level comparison here. We find that large mismatch exists between these two lists. Specifically, 59\% of developers on the two rankings are different, and those appearing in both rankings correspond with different orders as well. This indicates that existing code commit metric might significantly underestimate the influences and contributions of many developers from issue discussion aspects. Besides, although the number of comments can represent the frequency of developer' participation in issue discussion to a certain extent, it misses abundant semantic details reflecting underlying leadership behaviors, as illustrated in the examples above. Therefore, this study is motivated to develop an effective approach to mine and identify such communication associated leadership, to offer a complimentary perspective to existing studies.

As an example, an OSS community can integrate our leadership identification tool with their issue tracking system, to monitor different leadership behaviors at a particular frequency. The metrics data can enable informed decisions such as understanding leadership dynamics and influences, recommending leadership styles for stimulating issue discussion, reflecting on leadership variation, etc.

\section{Categorizing Leadership Behaviors}
\label{sec:leadership}

This section presents a set of six categories of ``emergent leadership'' behaviors in collaborative issue discussion processes in OSS communities. Unlike previous studies on OSS leadership, we take a different approach in this study, by considering leadership as an influence process, not the totality of the behaviors of a particular developer. To that end, we aim at examining leadership behavior at individual comment level, instead of at the developer level (e.g. measuring the corpus of comments by a person). Such a conceptualization of leadership has two main advantages. First, it allows us to observe comment-level leadership behavior with finer granularity. Specifically, this enables us to examine immediate impact of each leadership behavior on other developers (e.g., as demonstrated by responding comments from others) and contextual factors surrounding leadership influence process, such as stimulating downstream discussions, and/or accelerating issue resolution. Second, using these comment-level leadership behaviors would also allow us to aggregate behaviors to the individual level. As OSS communities are constantly changing, more research is needed for aggregating comment-level metrics at the individual/team level, with respect to particular usage scenarios.

Existing leadership literature has offered many taxonomies of leadership behaviors.  We build the six categories based on reviewing transformational leadership \cite{bass1996multifactor}, Yukl’s comprehensive taxonomy \cite{yukl2002hierarchical,yukl2012effective}, shared leadership\cite{Li2012leadership}, and emergent leadership\cite{ferebee2012emergent}. The six leadership categories were adapted from 12 Yukl’s comprehensive taxonomy \cite{yukl2002hierarchical,yukl2012effective} according to the matching leadership functions in the OSS context, particularly for achieving the goal of efficient issue resolution. Table \ref{tab:cmt_types} summarizes these leadership categories with description and comment examples, along with the mapping with the 6 selected Yukl’s categories \cite{yukl2002hierarchical,yukl2012effective}. The other 6 Yukl's categories concerns more on change-oriented, hierarchy and external components of leadership, which are not included in this study due to less relevance to the OSS issue discussion context. More specifically, the leadership included in this study is as follows:

\begin{table*}[!t]
\caption{Categories of emergent leadership behaviors in OSS projects}
\label{tab:cmt_types}
\begin{center}
\scalebox{0.8}{
\begin{tabular}{l|l|l|l}
\hline
Comment Type & Description & Examples & Yukl's Mapping \\ 
\hline
LD1 (Proposal) & \tabincell{l}{Propose a solution or alternative} & \tabincell{l}{1. Try install Bitcoin Core 0.17. \\ 2. You could try to remove Homebrew's binutils.} & Planning \\ 
\hline
LD2 (Redirection) & \tabincell{l}{Guide to other topics or places to discuss} & \tabincell{l}{1. This is a duplicate of \#17576. \\ 2. This should be an issue in ... repo.} & Supporting \\ 
\hline
LD3 (Confirmation) & \tabincell{l}{Confirm an issue or opinion } & \tabincell{l}{1. I agree with @sipa. \\ 2. Confirming that I got the same error on ...} & Recognizing \\ 
\hline
LD4 (Inquiry) & \tabincell{l}{Ask for more information} & \tabincell{l}{1.  Where did you get the code from? \\ 2. What version of Boost are you using?} & Consulting \\ 
\hline
LD5 (Operation) & \tabincell{l}{Suggest to close or reopen the issue report} & \tabincell{l}{1. Going to close due to lack of information. \\ 2. It's just a workaround... plz reopen this.} & Monitoring operation \\ 
\hline
LD6 (Volunteer) & \tabincell{l}{Voluntary acceptance of a task} & \tabincell{l}{1. I would like to work on this. \\ 2. I will open a PR fixing this soon.} & Clarifying roles \\ 
\hline
\end{tabular}  
}
\end{center}  
\end{table*}

\textbf{LD1 (Proposal).} This category corresponds to the short-term planning category from Yukl's taxonomy, and refers to the idea proposal type of issue discussion, such as proposing ideas for resolving an issue, suggesting unexplored processes and resources to be explored. Issue discussion is an intelligence-intensive problem solving process, and LD1 discussions usually shape potential issue resolution plans, and accelerate the issue resolution cycles. In addition, we believe it is also consistent with the intellectual stimulation/inspiration category in transformational leadership \cite{bass1996multifactor, neufeld2019leadership}.

\textbf{LD2 (Redirection).} This category corresponds to the supporting category from Yukl's taxonomy, and refers to comments providing supportive information to redirect attentions to a new topic (e.g., concepts, processes, or resources) or to a more relevant information sources. Due to the open and interconnected nature of OSS communities, such topics and information sources may be either internal (e.g., a related issue within the OSS project) or external (e.g., a site outside the OSS project). 

\textbf{LD3 (Confirmation).} This category corresponds to the recognizing category from Yukl's taxonomy, and refers to the leadership function in confirming or reiterating an idea or opinion, which facilitates the consensus processes among members, through implicit voting or shaping shared perception among developers. 

\textbf{LD4 (Inquiry).} This category corresponds to the consulting category from Yukl's taxonomy, and refers to cases where developers ask for more clarification question about the issue being reported/discussed. Example follow-up inquiries include asking for specific project version info or issue reproduce steps. Such issue comments typically provide some sort of confirmation to former opinions, and facilitate the issue discussion threads. 

\textbf{LD5 (Operation)} and \textbf{LD6 (Volunteer)} represent two distinct types of administrative functions during issue life cycle, corresponding to the monitoring and clarifying roles categories, respectively, from Yukl's taxonomy. LD5 refers to comments related to either closing or reopening an issue, and LD6 relates to volunteering in undertaking some wrap-up bug fixing actions, respectively. 

To validate the representativeness of the 6 comment-level leadership categories, we design and conduct a survey to gather feedback from OSS developers. The main question asked is whether they agree that these categories capture the relevant leadership behaviors which contribute to the issue discussion and resolution processes.
We distribute the survey to the contributors of 10 popular projects (as shown in Table \ref{tab:repos}) on GitHub, and received 27 responses\footnote{Due to space limit, we put the details of the survey on our project website.}. The results, as shown in Figure \ref{fig:survey}, indicates that an average of 71\% of developers believe that these six categories of leadership provide a good representation of emergent leadership behaviors in OSS issue discussion communications. Furthermore, even the most controversial LD5 has no strong disagree, and the number of disagrees is only 26\% (7/27). One possible reason for the disagreement on LD5 might be because developers tend to link leaders or leadership behaviors with more technical contribution, while LD5 mostly deals with process monitoring and technical focused. 

\section{Approach}
\label{sec:approach}

When developers explain something similar, they are likely to use recurrent textual expressions in their comments \cite{winograd1986understanding}.
Existing studies frequently rely on such expressions to extract linguistic patterns in order to facilitate context understanding and task automation \cite{DBLP:conf/sigsoft/Zhao0BSWMW20,DBLP:conf/icse/0008ZBPL19,DBLP:conf/kbse/ShiCWLB17,DBLP:conf/icsm/PanichellaSGVCG15, DBLP:conf/europlop/Silva17}. 
In this section, we present a novel approach, {\tool}, which employs a set of heuristic linguistic patterns to automatically mine developers' issue comments, and identify the emergent leadership behaviors (as described in Section \ref{sec:leadership}). 

\begin{figure*}[!t]
\centering
\includegraphics[width=15cm]{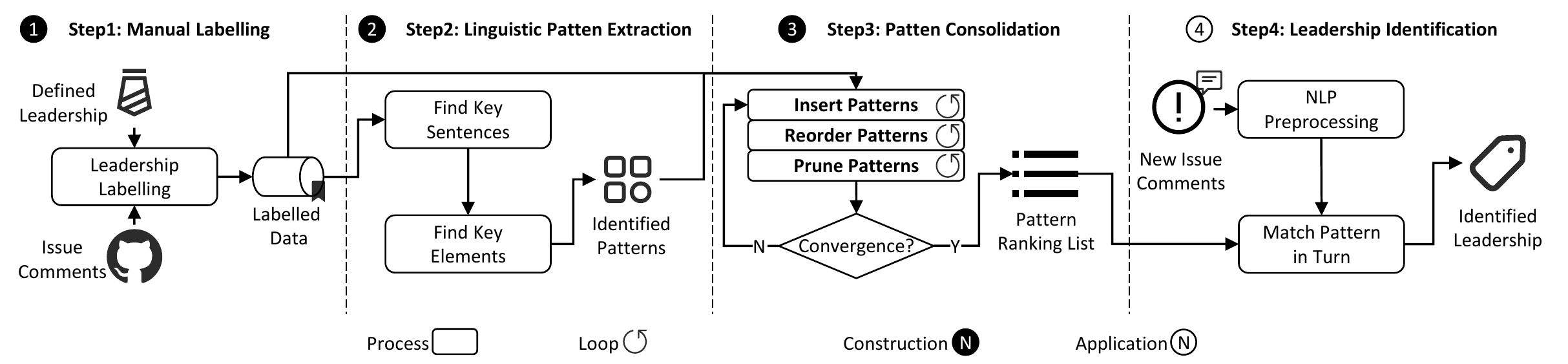}
\caption{{\tool} overview}
\label{fig:approach_overview}
\end{figure*}

Figure \ref{fig:approach_overview} illustrates the overview of the {\tool} approach,  consisting of four steps including: 1) Manual labelling of issue comments, in order to establish ground truth of leadership labels (Section \ref{subsec:manual_anno}); 2) Extracting linguistic patterns specifying the leadership behaviors from the comment in labelled data (Section \ref{subsec:HLP}); 3) Developing a pattern consolidating algorithm to reduce the penalty due to potential pattern over-fitting and over-loading (Section \ref{subsec:consolidating_patterns}); 4) For new issue comments, applying the pattern ranking list output from the previous step (\textit{Step 3}) to identify the leadership behaviors of the issue comments (Section \ref{subsec:leadership_identification}).

\subsection{Manual Labelling}
\label{subsec:manual_anno}
In this step, we will mark each issue comment with a corresponding leadership label (i.e., LD1, LD2, ..., or LD6), as introduced in Section \ref{sec:leadership}. Specifically, the first three authors individually label the same issue comments, then discuss and merge the results on each individual issue comment. The merged leadership labels serve as the ground truth in the later steps, i.e., pattern consolidation and evaluation.

To ensure the validity of the manual labelling outcomes, we adopt a labelling process in the existing work \cite{pustejovsky2012natural}, and the process includes four activities, i.e., defining a labelling schema, independently labelling by authors, group reviewing of individual labels, and building a final consensus corpus. 

Specifically, the labelling schema is based on the six categories of emergent leadership behaviors, as mentioned in Section \ref{sec:leadership}, plus an additional category (N) for comments not containing any leadership behaviors. A group discussion is conducted to ensure that everyone involved in labelling have a common understanding of these six leadership behaviors in open source issue discussion context. 

Then, each comment will be independently labelled by the first three authors and produce the label of a corresponding leadership behavior, or \textit{N} indicating no leadership behavior. The average Cohen’s Kappa is 0.75 between each pair of annotators, indicating substantial agreement among the annotators.

Next, a group meeting is held to review individual labelling results. When a discrepancy arises, each author involved in labelling provides rationale for his/her choice, and the group then discusses and reaches an agreement.  

Finally, after several rounds of discussions, agreements are reached for every comment and we get the final consensus corpus.

\subsection{Linguistic Pattern Extraction}
\label{subsec:HLP}

\begin{figure}[!t]
\centering
\includegraphics[width=10cm]{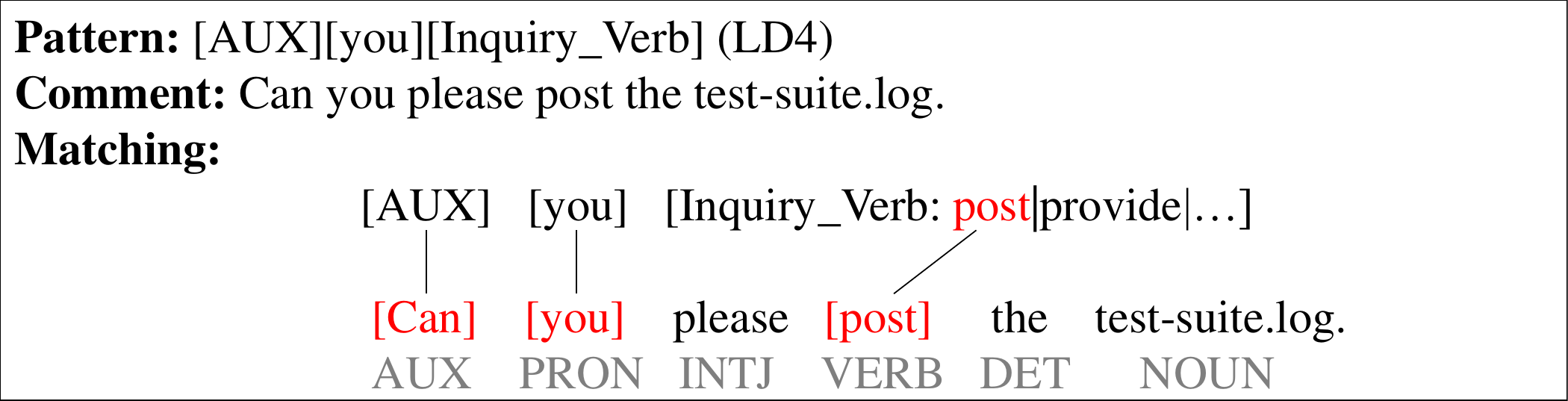}
\caption{Example of leadership identification with pattern}
\label{fig:match_example}
\end{figure}

The second step is to manually extract linguistic patterns from the textual content of issue comment, which can be used for predicting each of the six leadership behaviors. 
For all comments with leadership labels (i.e., LD1 to LD6), we first extract the key sentence in the comments which determines it belonging to certain leadership behavior. Take the following LD4 (Inquiry) comment as an example \textit{``Can you provide more information? I've just done a build of master using the 64 bit Windows instructions and it's working correctly.''} The key sentence is \textit{``Can you provide more information?''}, since the other sentences help little in identifying the leadership behaviors. After that, the more structured part of the sentence like \textit{``...can...you...provide...''} will be identified and the linguistic pattern will be generated base on it. In this example, a pattern containing three elements about \textit{``can''}, \textit{``you''} and \textit{``provide''} will be extracted.

The patterns generated from the above process is relatively large, if only concrete word/phrase elements are considered. For example, \textit{``Can you provide more information''}, \textit{``Could you give us the steps to reproduce''}, \textit{``Would you mind uploading the dockerfile''} and \textit{``Are you sure you're not running master''} all have similar structures and belong to LD4 (Inquiry). Since the words are different, we get four patterns: \textit{``can you provide''}, \textit{``could you give''}, \textit{``would you mind uploading''} and \textit{``are you sure''}. This will lead to a huge pattern set, which is difficult to maintain.

In order to reduce the number of patterns, we follow the existing research \cite{DBLP:conf/icse/0008ZBPL19,DBLP:conf/kbse/ShiCWLB17} by considering not only words as elements of a patterns, but also the Part-of-Speech(POS) tag of the words, and use dictionary to represent a group of words that have similar usage. For a pattern with \textit{Dictionary}, it will be used as a template which means each word in the dictionary can be applied to the corresponding positions and formulates a specific pattern. As a result, the four LD4 patterns in the previous paragraph can be merge into \textit{``[AUX(POS)] [you(word)] [Inquiry\_Verb(Dictionary)]''}. This 
not only reduces the total number of patterns, but also enables more sophisticated pattern matching. Figure \ref{fig:match_example} shows how this pattern can be applied in leadership identification.

Two of the authors participate in the process and employ the open card sorting \cite{DBLP:journals/es/RuggM05} during this process. Specifically, we do not pre-define the relevant patterns and dictionaries. Each author extracts patterns independently on the collected data and organize their corresponding pattern set. Then group meetings are conducted to discuss and merge the pattern sets. After that, the merged pattern set will be maintained by the authors together and each author can add new patterns to the set or refine existing patterns in their subsequent pattern extraction. Finally, we get a set of patterns that can be used to match comment and identify leadership.
For a complete list of all linguistic patterns, please refer to our project website on GitHub, and a total of 304 patterns are extracted for the follow-up leadership identification (see details in Table \ref{tab:rq1_iter}).

\subsection{Pattern Consolidation}
\label{subsec:consolidating_patterns}
In this step, we develop an iterative pattern consolidation algorithm to merge  patterns extracted from individual projects, monitor and tune the performance change of {\tool} against the newly merged pattern set in each iteration. This step is essential due to two main reasons: (1) patterns extracted from individual projects are at the risk of pattern over-fitting; (2) in cases that multiple patterns can be simultaneously applicable, appropriate priorities need to be assigned in order to avoid complexity and ambiguity caused by overloaded patterns. To address the pattern over-fitting and overloading issues, we develop an algorithm to automate the pattern merge and consolidation process. This algorithm consists of three sequential tasks (i.e., insert, reorder, and prune), and stops when the performance of {\tool} converges to a satisfactory level. We will elaborate each of the three tasks next.

\subsubsection{\textbf{Insert Patterns}}
\label{subsubsec:insert_patterns}
This step takes two input: one is a target pattern ranking list (i.e., initially an empty list), and the other is a new pattern set to be inserted into the target pattern ranking list.

Specifically, for each new pattern, the algorithm will iteratively examine the impact of merging it into the target pattern ranking list, and only accept it when it positively contributes to the prediction performance of {\tool}. 
Following an existing study \cite{DBLP:conf/sigsoft/Zhao0BSWMW20,DBLP:conf/icse/0008ZBPL19,DBLP:conf/kbse/ShiCWLB17}, we choose the F1-Score (mentioned in \ref{subsec_exp_metric}) as the performance metric.
Specifically, the algorithm will probe every candidate position in the pattern ranking list for inserting a new pattern, calculate the performance of {\tool} in each probing, record the best performance and the corresponding optimal insertion position. As a result, a new pattern will only be inserted into the merged list if it positively contributes to the performance of {\tool}, and the optimal position is the one corresponds to the best performance. Otherwise, the new pattern will be discarded and will not be merged into the pattern ranking list.

\subsubsection{\textbf{Reorder Patterns}}
\label{subsubsec:update_ranking}

We observe that, after the previous merging process, some issue comments may be incorrectly identified according to a loosely relevant pattern with a higher priority, while a more relevant pattern is ranked lower in the list and never be applied.
To address this, the algorithm reorders the ranking of patterns by examining these incorrectly labelled comments, to promote the overall performance on the whole dataset. The ranking update is conducted iteratively and dynamically in terms of each incorrectly labelled comment by current pattern ranking list.

In detail, for an incorrectly labelled comment (the comment mislabelled by the current pattern ranking list), we define the highest ranking pattern which correctly labels it as \emph{realwinner}, while the patterns which incorrectly match it and higher than \emph{realwinner} as the \emph{fakewinners}. If \emph{fakewinners} list is not empty, algorithm will try the ranking update. Specifically, algorithm will insert the \emph{realwinner} ahead of the \emph{fakewinners} in the pattern ranking list, and examine the leadership labelling performance on the whole dataset (i.e., F1-Score in Section \ref{subsec_exp_metric}). If the performance increases with this new pattern ranking list, the update is accepted and the current pattern ranking list will change to the new pattern ranking list. Then the algorithm iterates with the next incorrectly labelled comment found by the current pattern ranking list until it remains unchanged. As a result, we obtain the updated pattern ranking list which will be used next.

\subsubsection{\textbf{Prune Patterns}}
\label{subsubsec:prune_patterns}
During this task, the algorithm will identify excessive patterns whose existance negatively contributes to the performance of {\tool}.
Similar to the previous task, the pattern pruning is conducted iteratively and dynamically by examining each incorrectly labelled comment.

In detail, for each incorrectly labelled comment, the algorithm first detects the \textit{realwinner} (i.e., the highestly ranked pattern which is able to correctly label the comment), as well as all \textit{fakewinners} (i.e., the patterns which incorrectly match the comment and higher than \textit{realwinner}).
After that, the algorithm will remove the \textit{fakewinners} in the pattern ranking list, and accept the update if the performance (i.e., F1-Score in Section \ref{subsec_exp_metric}) increases.
Then the algorithm iterates with the next incorrectly labelled comment found by the refined pattern ranking list until it remains unchanged.

After that, we will obtain the pattern ranking list and if the patterns from a new project need to be added, the algorithm will start executing from \textit{Insert Patterns} again and use this list as the input (instead of the original empty list).

\subsection{Leadership Identification}
\label{subsec:leadership_identification}

After completing the construction of pattern ranking list, we can use it to identify leadership behaviors. Specifically, for a new issue comment, {\tool} first preprocesses it to get the words and the Part-of-Speech (POS) tags.
Then, it matches the preprocessed data with element(s) specified in a given pattern. We employ the following two distance constraints between elements, adapted from existing study \cite{DBLP:conf/cikm/ShenSYC06}, to promote the performance of leadership identification.
The first constraint is that the latter matched element cannot appear before the prior matched element, since the pattern is extracted by traversing the words in sequence.
The second constraint is that the distance between two matched elements should be no more than three, which is to avoid the mismatch of patterns since the leadership behaviors are usually expressed in a concise way. 
{\tool} will automatically match the comment with the patterns in the list in order, and following existing study \cite{DBLP:conf/kbse/ShiCWLB17}, we choose the highest priority strategy which means the leadership of the comment is identified by the first matched pattern. 

\section{Experiment Design}
\label{sec_experiment}
\subsection{Research Questions}
\label{subsec_exp_RQ}
To evaluate the proposed approach, we answer the following three research questions:
\begin{itemize}
\item \textbf{RQ1 (Performance Convergence)} When can the generated linguistic patterns reach the state of convergence? 
\item \textbf{RQ2 (Performance Evaluation)} How effective is {\tool} in leadership identification on new projects? 
\item \textbf{RQ3 (Baseline Comparison)} Can {\tool} outperform other techniques in leadership identification? 
\end{itemize}

\subsection{Subject Projects}
\label{subsec_exp_projects}
We collect 10 popular open source software projects from GitHub for evaluation. The projects are selected with the following criteria: (1) Popularity: with 10k+ stars and 1.5K+ forks; (2) Activeness: regularly updated within 1 month of the data collection date; (3) Diversity: representing different OSS organizational structures from various domains. A brief summary of the experimental projects are summarized in Table \ref{tab:repos}. 

For the experimental projects, we use the GitHub's REST APIs to crawl the issues and related comments\footnote{The data is crawled between July 2020 and February 2021.}. We only crawl the comments of closed issues since they represent the complete issue life-cycle. In addition, since it is our interest to compare the emergent leadership behaviors with traditional leadership indicators, we also extract two indicators at the developer level: the number of commits and the number of followers. For each project, we randomly sample a portion of issues and obtain 1,000 related comments (in Table \ref{tab:repos}), and conduct manual labelling, as introduced in Section \ref{subsec:manual_anno}. The overview for the ground truth of the leadership labels is shown in Table \ref{tab:repos}. 

\begin{table*}[!t]
\caption{Subject projects}
\label{tab:repos}
\begin{center}  
\scalebox{0.88}{
\begin{tabular}{l|l|l|l|l|l|l|l|l|l|l|l|l|l}
\hline
\multicolumn{7}{c|}{Basic information} & \multicolumn{7}{c}{Leadership labels} \\
\hline
ID & Project & Link & Domain & Stars & \#Comments & \#Issues & LD1 & LD2 & LD3 & LD4 & LD5 & LD6 & TotalLD\\ 
\hline
P1 & bitcoin & \cite{bitcoin} & digital currency & 57k & 28,504 & 4,997 & 95 & 98 & 123 & 132 & 55 & 18 & 521\\
\hline
P2 & sklearn & \cite{sklearn} & machine learning & 47k & 40,432 & 6,287 & 81 & 47 & 63 & 73 & 105 & 81 & 450\\
\hline
P3 & ember.js & \cite{emberjs} & js framework & 22k & 32,225 & 5,833 & 69 & 54 & 132 & 77 & 89 & 43 & 464\\
\hline
P4 & brew & \cite{brew} & package manager & 29k & 14,269 & 2,694 & 74 & 55 & 89 & 85 & 34 & 38 & 375\\
\hline
P5 & atom & \cite{atom} & text editor & 56k & 83,004 & 15,621 & 50 & 255 & 141 & 42 & 56 & 6 & 550\\
\hline
P6 & bokeh & \cite{bokeh} & visualization & 15k & 21,432 & 4,573 & 74 & 76 & 79 & 65 & 60 & 56 & 410\\
\hline
P7 & efcore & \cite{efcore} & entity framework & 11k & 55,820 & 11,596 & 87 & 140 & 47 & 162 & 72 & 7 & 515\\
\hline
P8 & knex & \cite{knex} & SQL query & 15k & 10,337 & 2,177 & 166 & 28 & 87 & 104 & 56 & 39 & 480\\
\hline
P9 & roslyn & \cite{roslyn} & compiler & 15k & 76,862 & 14,722 & 61 & 123 & 74 & 56 & 60 & 37 & 411\\
\hline
P10 & solidity & \cite{solidity} & program language & 12k & 11,567 & 2,513 & 44 & 52 & 84 & 90 & 46 & 28 & 344\\
\hline
\end{tabular}
}
\end{center}
\centering
\end{table*}

\subsection{Data Pre-processing}
\label{subsec:preprocess}

We observe that some developers would refer to the previous comments in their own comments. The referenced descriptions may contain another developer's leadership, and should be removed to reduce potential noise leading to incorrect matching. In this study, we implement this via removing the node which is used to highlight the reference descriptions in the comment's HTML text. 

In addition, we use Stanford natural language processing tool Stanza \cite{qi2020stanza} to process the text in lowcase, tokenization, part of speech (POS) tagging and lemmatization, so as to obtain the data applied to the construction process of {\tool}.
We also use Urlextract\footnote{\url{https://github.com/lipoja/URLExtract}} to annotate the URL contained in the comments, which is the element in certain patterns.

\subsection{Evaluation Metrics}
\label{subsec_exp_metric}

We employ three commonly-used metrics to evaluate leadership identification performance, i.e., precision, recall and F1-Score. 

\textbf{Precision} measures how precise a classifier is in predicting the comments with specific category of leadership. 

\textbf{Recall} measures the ability of classifier to find the comments with specific category of leadership.

\textbf{F1-Score} is the harmonic mean of precision and recall. 

For the precision, recall, and F1-Score of multiple categories of leadership (as shown in Figure \ref{fig:model_perf}), we calculate these metrics of each category and use the average value.

\subsection{Experiment Setup}
\label{subsec:experimentsetup}
\subsubsection{\textbf{Experiment for answering RQ1}}

We start with the set of linguistic patterns established from one project, and then employ an iterative process as follows: 1) add another set of linguistic patterns established from a new project; 2) apply the pattern consolidation algorithm from Section \ref{subsec:consolidating_patterns} to generate a  pattern ranking list; 3) input the pattern ranking list to {\tool} for leadership identification; 4) evaluate the performance of {\tool} using the previously defined metrics. Throughout the iterative process, we continuously monitor the performance change of {\tool}, and stop the process when the performance converges (i.e., when the additional performance gain is less than 0.01). By the convergence time, we stop to incorporate new patterns, and finalize the pattern ranking list in {\tool} as the default pattern list used for further experiment and evaluation. 

\subsubsection{\textbf{Experiment for answering RQ2}} To evaluate the performance of {\tool}, we use the remaining projects (projects unused in RQ1) as test sets and measure the identification performance of {\tool} on each test project. In addition, we will also examine the performance of {\tool} across the six leadership categories to obtain more comprehensive performance results. 

\subsubsection{\textbf{Experiment for answering RQ3}} We employ commonly-used machine learning based and deep learning based text identification approaches as baselines to compare with {\tool}. 
Specifically, we use Term Frequency and Inverse Document Frequency (TF-IDF) \cite{DBLP:books/daglib/0021593} to vectorize textual descriptions of comments, and use the term vectors as input for identification. For selection of machine learning classifiers, we evaluate four commonly-used machine learning models: Logistic Regression (LR) \cite{berkson1944application}, Support Vector Machine (SVM) \cite{cortes1995support}, Decision Tree (DT) \cite{DBLP:books/wa/BreimanFOS84}, and Random Forest (RF) \cite{ho1995random}, and due to the limited space, we only show the best classifier (SVM). For the deep learning based classifiers, we utilize TextCNN \cite{DBLP:conf/emnlp/Kim14}, BiLSTM \cite{DBLP:conf/ijcai/LiuQH16}, CNN and RNN with Bert pre-training model \cite{DBLP:conf/naacl/DevlinCLT19}.

In addition, we further explore whether our extracted patterns are helpful to the machine/deep learning models, following the existing studies \cite{DBLP:conf/kbse/ShiCWLB17,DBLP:conf/icse/0008ZBPL19,DBLP:conf/sigsoft/Zhao0BSWMW20}.
In detail, we enrich the model inputs with the knowledge about pattern matching, which will be consider as a boolean vector, i.e., if a comment match with certain patterns, the corresponding position of the vector is set to 1, otherwise it is 0. We also run the above mentioned machine/deep learning models on the combined vectors, and they are short for <Model\_Name-P> (e.g., TextCNN-P). In total, we have 10 baselines for comparison. 

\section{Results and Analysis}
\label{sec_results}

\subsection{Answering RQ1. Performance Convergence}
\label{subsec_result_RQ1}
As introduced in Section \ref{subsec:experimentsetup}, We start with the pattern set extracted from one project, and iteratively add a pattern set extracted from a new project, consolidate into the pattern ranking list following the algorithm described in Section \ref{subsec:consolidating_patterns}.
In each iteration, we record the number of patterns added, deleted and changed, as well as the performance of {\tool}. We find that the performance of {\tool} converges after five iterations. Table \ref{tab:rq1_iter} summarizes the results from each of the five iterations. 

\begin{table}[!t]
\caption{Performance change with iterations (RQ1)}
\label{tab:rq1_iter}
\begin{center}  
\begin{tabular}{p{1.2cm}|p{1cm}|p{1cm}|p{1cm}|p{1cm}|p{1cm}}
\hline
\multicolumn{6}{c}{Pattern extraction results} \\
\hline
Projects & P1 & P1-P2 & P1-P3 & P1-P4 & P1-P5 \\ 
\hline
\#Patterns & 198 & 250 & 287 & 300 & 304 \\ 
\hline
\#Add & - & 55 & 41 & 13 & 4 \\ 
\hline
\#Delete & - & 2 & 4 & 0 & 0 \\ 
\hline
\#Change & 3 & 5 & 1 & 0 & 0 \\ 
\hline
\multicolumn{6}{c}{Leadership identification performance} \\
\hline
Precision & 0.81 & 0.82 & 0.83 & 0.83 & 0.83\\
\hline
Recall & 0.74 & 0.82 & 0.84 & 0.85 & 0.85\\
\hline
F1-Score & 0.76 & 0.82 & 0.83 & 0.84 & 0.84\\
\hline
\end{tabular}  
\end{center}  
\end{table}

Across the five iterations, we can see that the number of patterns increase from 198 to 304, and the accuracy increases from 0.81 to 0.85. We also observe that different projects share common linguistic patterns in issue comments, as seen by the smaller number of newly added patterns in each iteration. In addition, the most pattern changes caused by newly added patterns is in the second iteration, when adding the pattern set of \textit{sklearn}, corresponding to 55 added, 2 deleted, and 4 changed patterns. Though there are new patterns added in the later two iterations as well, however, their added impact on the performance is diminishing. When the fifth project is added, only very slight changes are observed in the pattern list and the three metrics. This means that the information from the fifth project does not bring much contribution to the pattern extraction, consolidation and performance increase. We believe that the results can support the conclusion on the sufficient state of convergence. 

In the remaining experiments, the set of patterns and their rankings obtained from the fifth iteration are finalized in {\tool}, and used to investigate other research questions.

\subsection{Answering RQ2. Performance Evaluation}

\begin{table}[!t]
\centering
\caption{Performance on unfitted projects (RQ2)}
\label{tab:rq2_unfitted}
\begin{tabular}{l|l|l|l|l|l|l}
\hline
Projects & bokeh & efcore & knex & roslyn & solidity & \textbf{AVG} \\
\hline
Precision & 0.80 & 0.83 & 0.82 & 0.82 & 0.84 & 0.82\\
\hline
Recall & 0.77 & 0.84 & 0.77 & 0.81 & 0.78 & 0.79\\
\hline
F1-Score & 0.77 & 0.83 & 0.79 & 0.81 & 0.80 & 0.80\\
\hline
\end{tabular}%
\end{table}%

\begin{figure}[!t]
\centering

\subfigure[Precision]{
\includegraphics[scale=0.6]{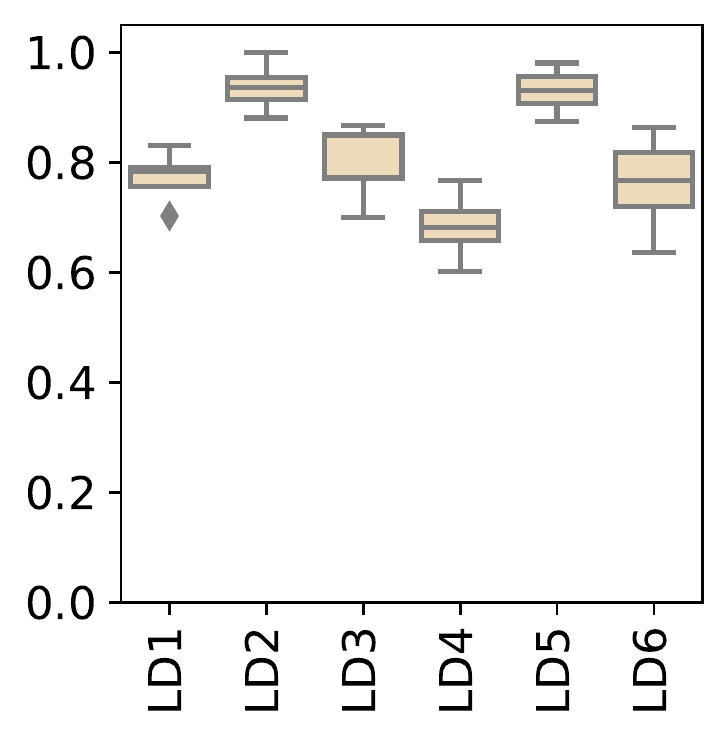}
}\hspace{-3mm}
\subfigure[Recall]{
\includegraphics[scale=0.6]{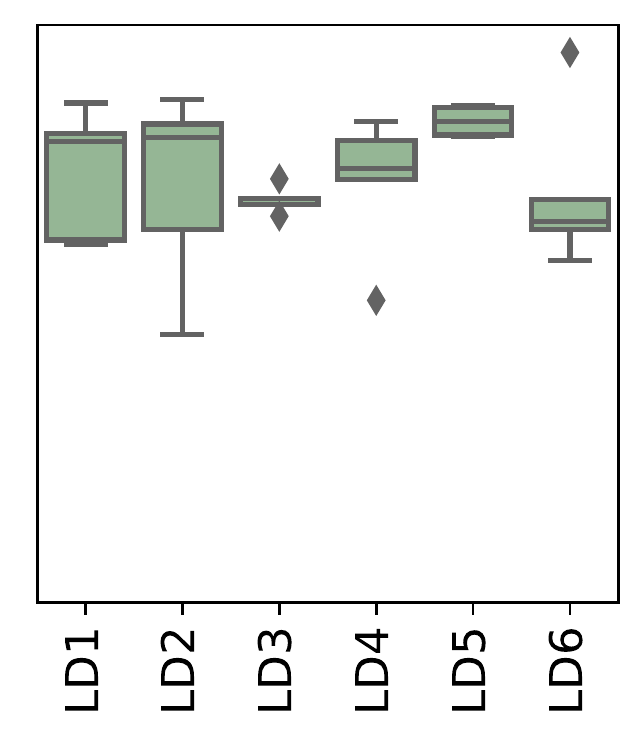}
}\hspace{-3mm}
\subfigure[F1-Score]{
\includegraphics[scale=0.6]{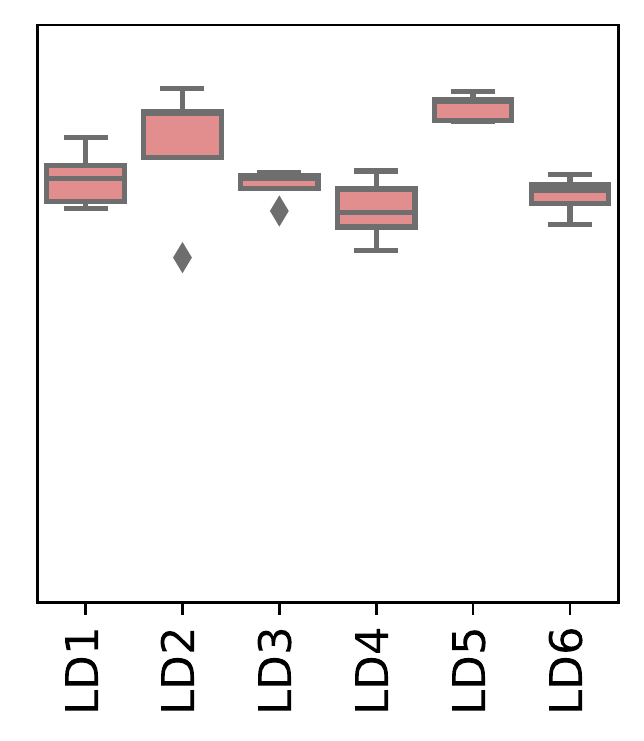}
}

\caption{Performance on six leadership categories (RQ2)}
\label{fig:cls_result}
\end{figure}

Table \ref{tab:rq2_unfitted} presents the performance of {\tool} applying to the five new  projects. 
The results show that {\tool} has an average precision of 82\%, and an average recall of 79\% across the five projects. In addition, the results are quite close to those from Table \ref{tab:rq1_iter}, which indicates the performance stability of {\tool}, as well as the generality of the pattern set and their relative ranking in {\tool}. It is also noticeable that the performance is slightly higher than the inter-rater agreement level of 0.75 during the manual annotation process (Section \ref{subsec:manual_anno}). A possible reason is associated with the group consensus based on majority vote. More specifically, some inter-raters discrepancy involves long comments containing multiple sentences, in which two or more sentences insides a long issue comment may correspond to different leadership labels. We employed group discussion and majority vote to keep 1 dominant leadership label. This handling might lead to the performance of {\tool} slightly higher than the interrater’s initial agreement level. Nevertheless, we observe a slight decrease in precision and recall, compared with the performance in Table \ref{tab:rq1_iter}. One possible reason is that these five new projects might contain new patterns in expressing leadership behaviors, which necessitates further tuning or addition investigation to include more patterns to improve the performance.

We also examine the performance of {\tool} in identifying each leadership category on the five new projects, as shown in Figure \ref{fig:cls_result}. From the comparison, we can see that LD2 (Redirection) and LD5 (Operation) are among the categories with highest performance, meaning relatively easier to identify. 
These two categories both reach a median precision over 0.9, median recall over 0.8. The lowest performance is observed in LD4 (Inquiry) and LD6 (Volunteer), yet with a satisfactory median of 0.70 F1-Score. 

\subsection{Answering RQ3. Baseline Comparison}

\begin{figure}[!t]
\centering

\subfigure[Precision]{
\includegraphics[scale=0.6]{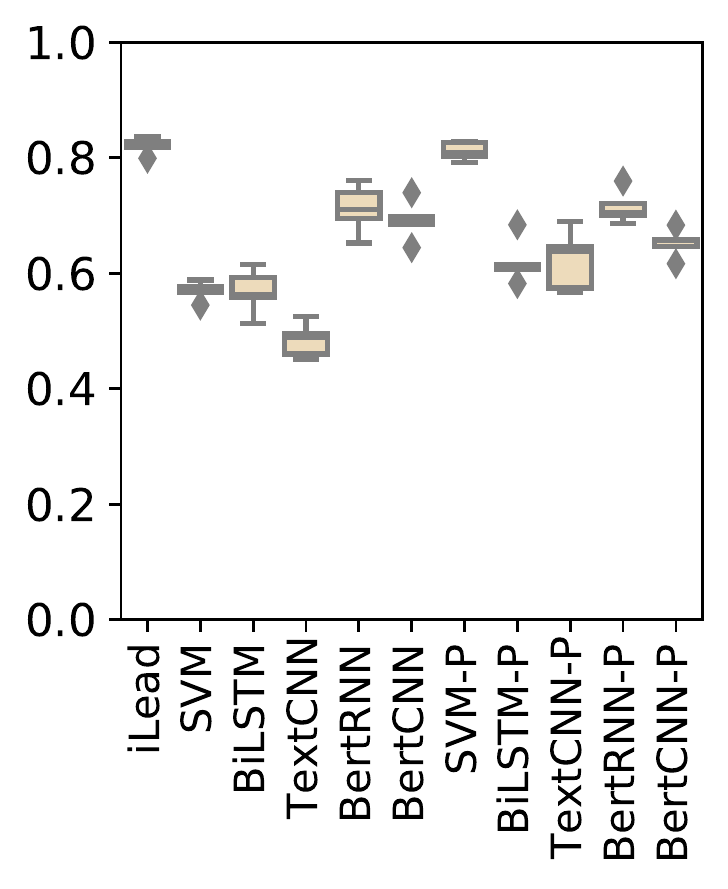}
}\hspace{-3mm}
\subfigure[Recall]{
\includegraphics[scale=0.6]{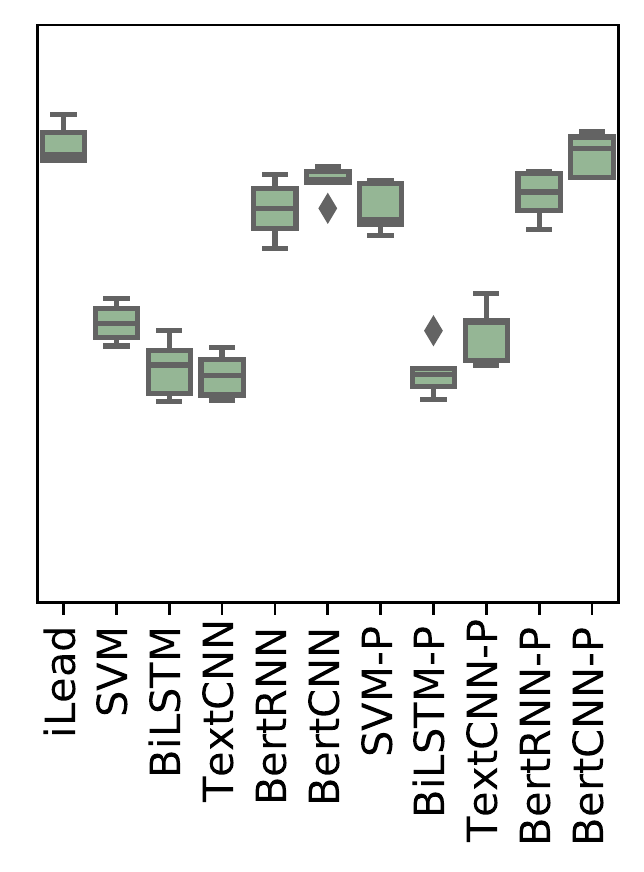}
}\hspace{-3mm}
\subfigure[F1-Score]{
\includegraphics[scale=0.6]{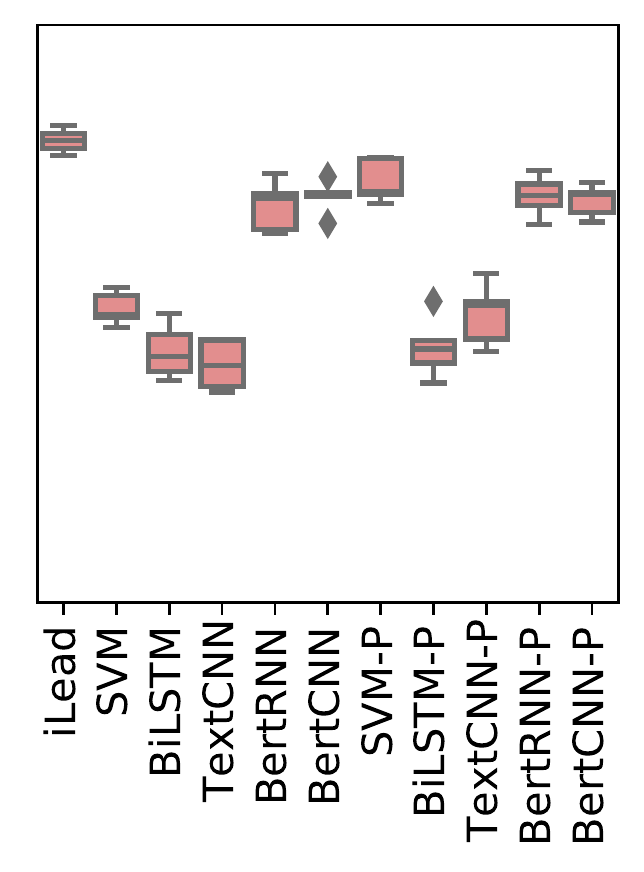}
}

\caption{Performance comparison with baselines (RQ3)}
\label{fig:model_perf}
\end{figure}

Figure \ref{fig:model_perf} demonstrates the comparison of the performance of {\tool} with the ten baselines. We can see that {\tool} outperforms all the baselines in all the investigated metrics, e.g., the median F1-Score of {\tool} is 13\% (i.e., (0.80-0.71)/0.71) higher than the best baseline \textit{SVM-P}. Furthermore, we conduct Mann-Whitney U test between the performance of {\tool} and the baselines. Results show that {\tool} significantly (p-value \textless 0.01) outperform all the baselines in F1-Score, and significantly outperform most of the baselines in precision (except SVM-P) and recall (except BertCNN-P).
This implies the effectiveness of our extracted patterns as well as the pattern consolidation algorithm in {\tool}. 

Among the baselines, the machine learning classifier with combined features, e.g., \textit{SVM-P}, is better than its counterpart only with textual features. For example, the median F1-Score achieved by \textit{SVM-P} is 42\% (i.e., (0.71-0.50)/0.50) higher than its counterpart \textit{SVM}. This implies our designed patterns are useful even when integrated into the machine learning models. 

Nevertheless, they still underperform our proposed approach, because our proposed approach has well-designed mechanism to optimize the patterns and their rankings, while the machine learners consider less of the application order of the patterns. We also observe that the deep learning based approaches do not achieve a satisfactory result as anticipated. This might because training an efficient deep learning models requires a large amount of labelled data, yet the dataset utilized in our experiment is far from that amount. Besides, there are many low frequency words which distinguish a certain category of leadership behaviors from others. But the deep learning models are not good at capturing these low frequency words.
\section{Revisiting OSS Leadership Landscape}
\label{sec:discussion}

This section describes a large-scale empirical evaluation with {\tool} on OSS projects. We apply {\tool} on all issue comments of the 5 fitted projects (from which the patterns are extracted) as shown in Table \ref{tab:repos}, to automatically identify the leadership behaviors, and explore: 1) the distribution of leadership behaviors; 2) the correlation between our emergent leadership behaviors and existing leadership indicators. 
We also conduct a human evaluation to verify the usefulness of {\tool}.

\subsection{Distribution of Leadership Behaviors}

\begin{figure*}[!t]
  \begin{minipage}[t]{0.48\linewidth}
    \centering
    \includegraphics[height=3.8cm]{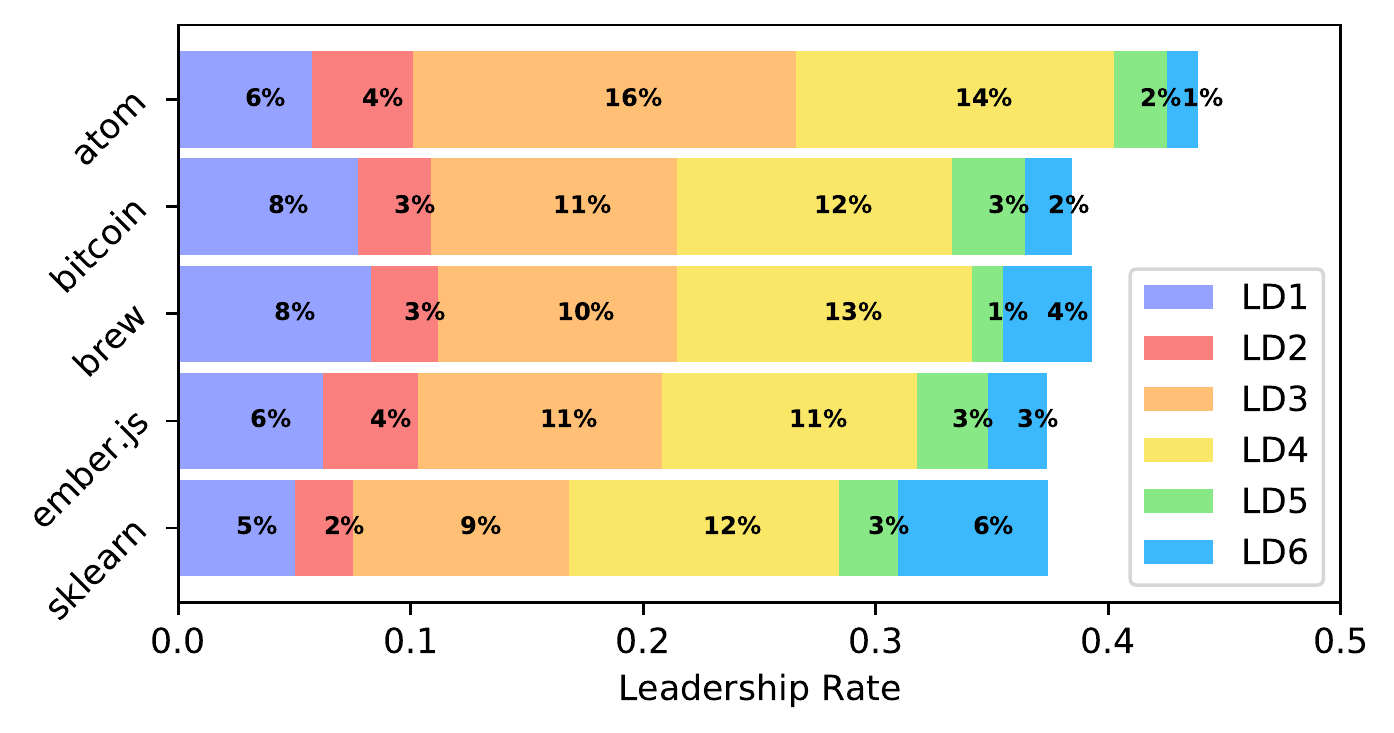}
    \caption{Distribution of leadership}
    \label{fig:ld_rate}
  \end{minipage}
  \begin{minipage}[t]{0.48\linewidth}
    \centering
    \includegraphics[height=3.8cm]{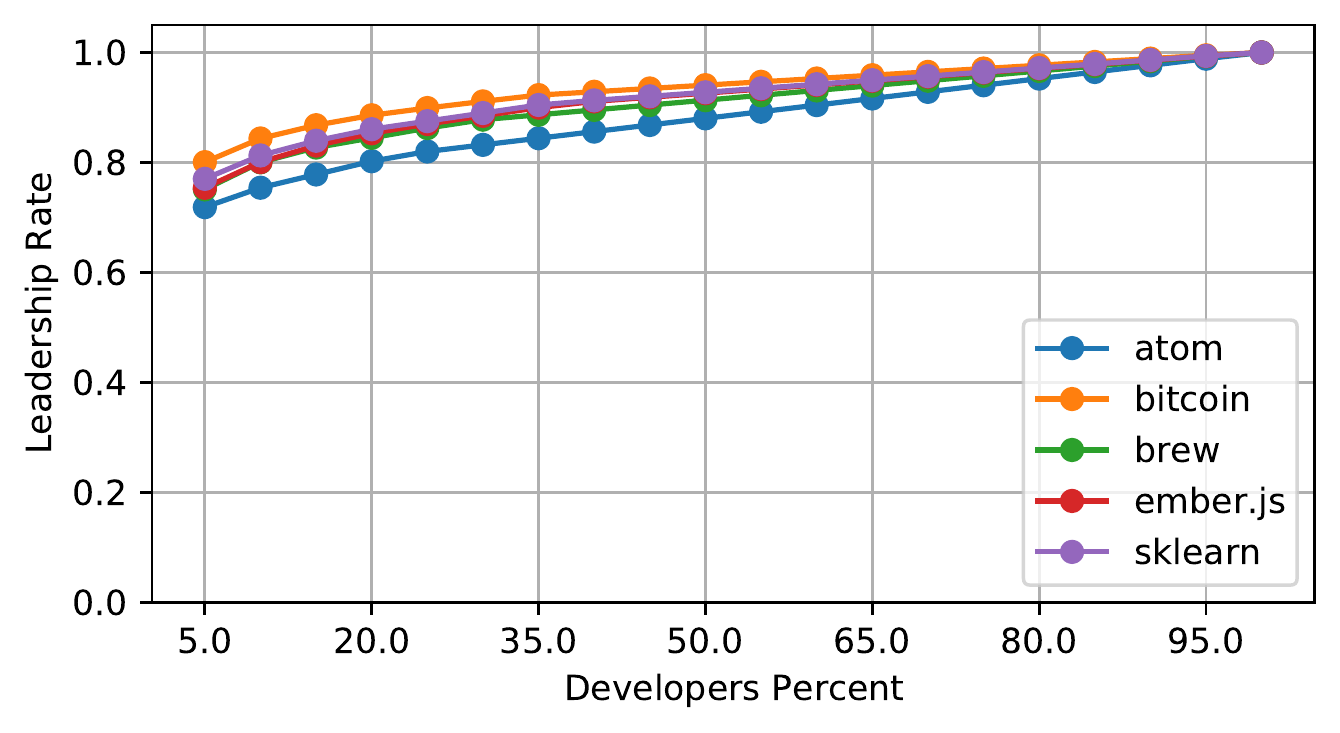}
    \caption{Cumulative distribution of developers' leadership}
    \label{fig:pareto}
  \end{minipage}
\end{figure*}

Figure \ref{fig:ld_rate} illustrates the distribution of the identified emergent leadership behaviors on the five projects. It is observed that: (1) Overall, leadership-associated comments account for 39\% of all comments on average (ranging from 37\% to 44\%); (2) The Top-3 dominant leadership categories are LD1 (Proposal), LD3 (Confirmation), and LD4 (Inquiry), corresponding to an average of 7\%, 11\%, and 12\% of all comments, respectively; and (3) The other three types, i.e., LD2 (Redirection), LD5 (Operation) and LD6 (Volunteer),  are observed at less frequency, corresponding to an average of 3\%, 2\%, and 3\% of all comments, respectively. These results provide quantitative evidences on developers leadership contribution in issue discussion, i.e., through brainstorming ideas, information inquiry, and consensus-based decision making. 

Figure \ref{fig:pareto} illustrates the pareto curves of individual developers' leadership-related actions in issue discussion, across the five projects. 
Two outstanding results are noted: (1) On the one hand, an average of 11\% (ranging from 5\% to 20\%) developers contribute to more than 80\% of leadership behaviors; (2) On the other hand, the majority (i.e., 61\%) developers (ranging from 40\% to 75\%) are associated with merely less than 10\% of leadership behaviors. This implies the potential need for more research to study OSS leadership through this new lens, e.g., supporting OSS leadership evolution and individual/team leadership coaching. 

\begin{figure}[!t]
\centering

\subfigure[Leadership Counts vs. Total Comments]{
\label{subfig:num_comments}
\includegraphics[scale=0.7]{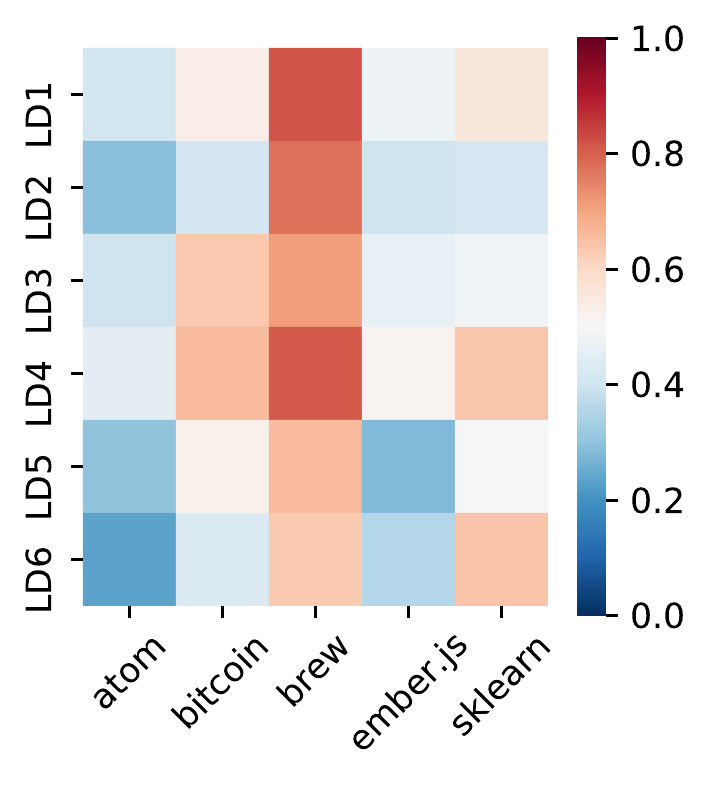}
}
\subfigure[Leadership\% vs. Total Comments]{
\label{subfig:rate_comments}
\includegraphics[scale=0.7]{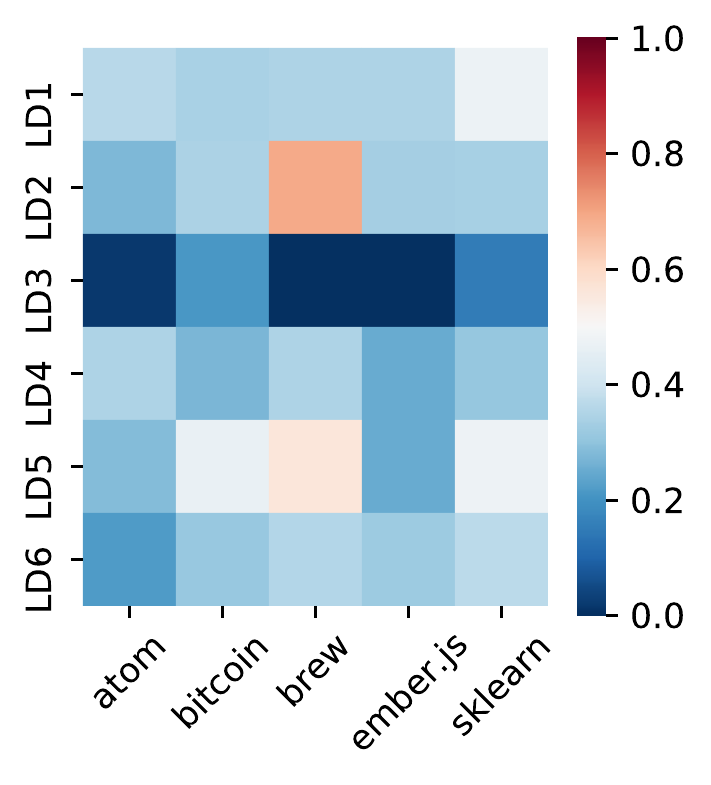}
}

\caption{Correlation between leadership indicators and total comments at individual developer level}
\label{fig:cmt_correlation}
\end{figure}

It is conceivable that leadership indicators may be correlated by the number of comments, corresponding to the common perception that when a developer publishes more comments, they are practicing more leadership. Indeed, we can observe the Top-11\% highly rated developers contribute 80\%+ leadership behaviors on average, as shown in Fig. 8. Particularly, for the Top-5\%, the correlation with the comment counts is 0.89. However, the focus of this study is be able to identify and cultivate emergent leadership behaviors among the majority middle-level developers to become more effective contributors and leaders. To explore whether the proposed emergent leadership indicators can offer more insights on such middle-level developers, we extracted two emergent leadership indicators for the middle 60\% of developers ranked by the total number of comments. The two emergent leadership indicators are the leadership count and the leadership percentage (i.e., leadership count/total comments) across all six leadership categories (i.e., LD1~LD6). Figure \ref{fig:cmt_correlation} shows the correlation results between the two new indicators of leadership and the traditional leadership metric of total comments at individual developer level, using data from 5 OSS projects. From Figure \ref{subfig:num_comments}, the high correlation is only observed on the \textit{brew} project, but not on the other 4 projects. 

As shown in Figure \ref{subfig:rate_comments}, the correlation between the percentage of leadership comments and the total comments is much lower, suggesting that developers making the same amount of comment contribution might correspond to different degree of emergent leadership behaviors. It can be concluded that from both charts, the number of comments does not necessarily mean more leadership, which indicates that our proposed leadership indicators provide a novel lens to offer new insights to study OSS leadership.

\subsection{Comparison among Leadership Indicators}

\label{subsec:ld_impact}
{\tool} is committed to identify the emergent leadership of developers and complement the existing viewpoints of developer contributions. To that end, we conduct a comparison analysis to measure OSS developers using different leadership metrics. For each developer, we aggregate the number of his/her comments corresponding with a specific leadership category. 
Then we conduct a correlation analysis to compare individual developers' emergent leadership behaviors with two existing leadership indicators, i.e. code contribution and community followers. 

\begin{figure}[!t]
\centering

\subfigure[\#Commits]{
\label{subfig:num_commits}
\includegraphics[scale=0.7]{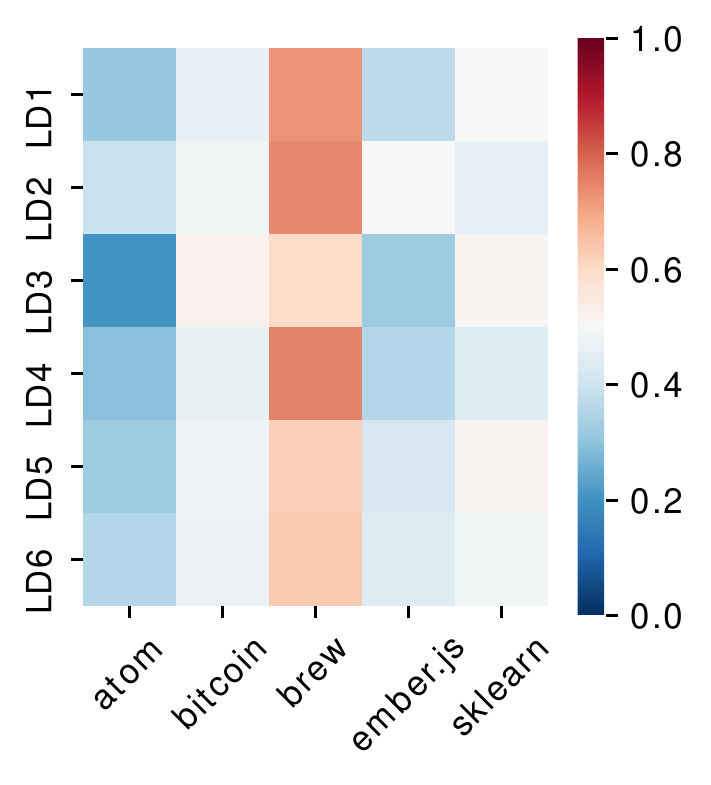}
}
\subfigure[\#Followers]{
\label{subfig:num_followers}
\includegraphics[scale=0.7]{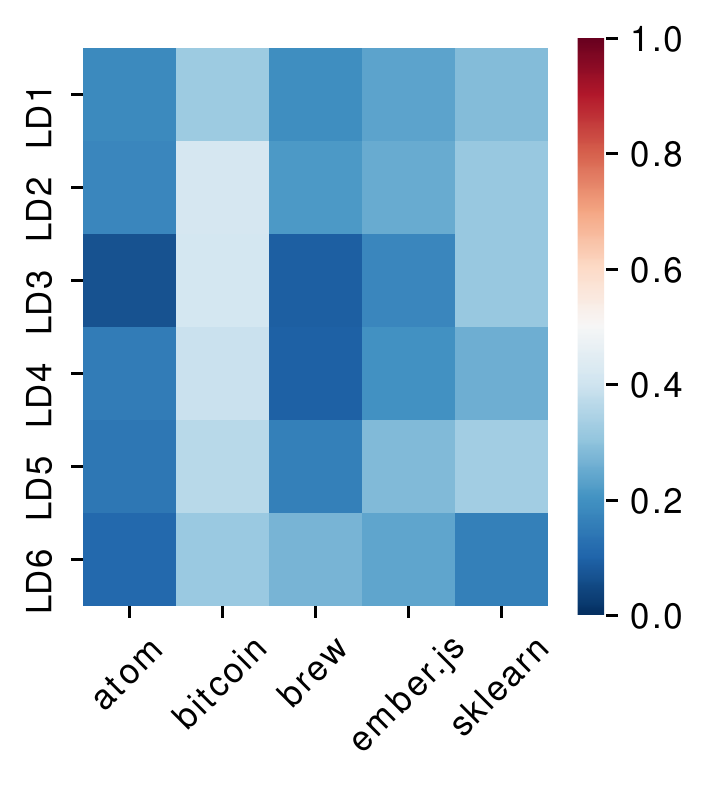}
}

\caption{Correlation between existing contribution indicators and \#leadership}
\label{fig:correlation}
\end{figure}

Figure \ref{subfig:num_commits} shows the correlation between the number of code commits and the number of identified leadership behaviors across the six emergent leadership categories. 
In general, the correlation between the number of leadership behaviors and the number of commits is low. This again confirms that emergent leaders are not always the leading code contributors. 
We can observe a relatively higher correlation between these two indicators in the \textit{brew} project. This may be because the \textit{brew} project entrusts its daily management activities to a third-party service organization\cite{sfconservancy}, and the developers focus more on the code contribution and issue discussion, i.e., resulting a higher correlation between these two indicators. Moreover, Figure \ref{subfig:num_followers} shows a much lower correlation between the number of followers and the number of leadership behaviors than that from Figure \ref{subfig:num_commits}. This confirms one of frequent findings in existing leadership literature \cite{kean2011followers, kelley1988praise}, which refers that leaders are not depending on how much followers they have, but how many leaders they create. In OSS community, there are a variety of motivating factors for following or not following, such as user's interest in learning, socializing,  obtaining updates, or easy access to others, etc. \cite{blincoe2016understanding}.

We think there are a couple of reasons for such low correlation. First, OSS developers may have different roles and preferences in different projects. For example, some prefer coding, while others participate more in the discussion. This suggests the need for a more dynamic and diverse lens to measure OSS leadership. Second, this reveals further opportunities for in-depth analysis to demonstrate the correlation between emergent leadership behaviors with existing leadership indicators, e.g., through data segmentation, time-lag analysis, etc. The low correlation between the identified emergent leadership behaviors and existing leadership indicators suggest that the identification of emergent leadership helps to broaden existing code contribution and followers viewpoints to embrace and recognize more comprehensive perspectives. 

\subsection{Leadership Influence}

As discussed earlier, we consider leadership as an influence process and measure leadership at individual comment level. Different leadership behaviors might have different impacts and the most obvious effect is on other discussions. To explore the influence of each leadership, we extracted 10 features encompassing two categories for each issue from the 5 projects. The first category focuses on examining the influence on process stimulation, and the second category focuses on examining issue resolution efficiency. We set an effect window of 24 hours, to extract process stimulation features reflecting later issue discussions after a leadership comment is posted. 

\begin{itemize}
\item other\_commenter: The number of commenter who is not the issue reporter or the author of this comment.
\item comment\_num: The number of comments.
\item reporter\_response: The number of comments posted by the issue reporter.
\item self\_response: The number of comments posted by the author of this comment.
\item other\_response: The number of comments posted by other commenters.
\item ld\_num: The number of leadership comments.
\item ld\_types: The number of leadership types.
\item word\_divergence: set(words) / len(words)
\item time\_from\_start: Time between the comment and the the opening of issue.
\item time\_to\_close: Time between comment and issue closing.
\end{itemize}

Table \ref{tab:ld_influence} shows the hypothesis testing results of the mean features' values of all 6 leadership indicators (Compared with non-leadership) at issue level, using Mann-Whitney U test. We use the positive signs to represent positive/increasing influences and use the negative signs to represent the negative/decreasing influences. Consistent results are observed across the 5 projects, for space consideration, Table \ref{tab:ld_influence} only shows the results on \textit{ember.js}. Results from other projects can be found on the study website \footnote{\url{https://github.com/20210827/iLead}}.

\begin{table}[!t]
\caption{Leadership hypothesis testing results}
\label{tab:ld_influence}
\begin{center}  
\scalebox{1}{
\begin{tabular}{l|l|l|l|l|l|l|l}
\hline
Feature & Category & LD1 & LD2 & LD3 & LD4 & LD5 & LD6 \\
\hline
other\_commenter & Process & (-)* & (-)*** & \# & (+)*** & (-)*** & \# \\
\hline
comment\_num & Process & (+)*** & (-)*** & \# & (+)*** & (-)*** & \# \\
\hline
reporter\_response & Process & (+)*** & (-)*** & (-)** & (+)*** & (-)*** & \# \\
\hline
self\_repsonse & Process & (+)*** & (-)*** & \# & (+)*** & (-)*** & \# \\
\hline
other\_response & Process & (-)* & (-)*** & \# & (+)*** & (-)*** & \# \\
\hline
ld\_num & Process & \# & (-)*** & (+)*** & (+)*** & (-)*** & \# \\
\hline
ld\_types & Process & \# & (-)*** & (+)*** & (+)*** & (-)*** & \# \\
\hline
word\_divergence & Process & (+)* & (-)*** & \# & (+)*** & (-)*** & \# \\
\hline
time\_from\_start & Resolution & (-)*** & (-)*** & \# & (+)** & (+)*** & (+)*** \\
\hline
time\_to\_close & Resolution & (-)*** & (-)*** & (+)*** & (+)*** & (-)*** & (+)* \\
\hline
\end{tabular}
}
\end{center}
\centering
\footnotesize{\#p >= 0.05, *p<0.05, **p<0.01, ***p<0.001}
\end{table}

We conclude: 
(1) LD1 (Proposal) tends to be able to trigger more discussions between the issue reporter and the LD1 publisher, and it may promote the issue to be solved;
(2) Comments after LD2 (Redirection) and LD5 (Operation) are less active. This is intuitive and indicates the issue originator's question is solved (LD2), and the issue has been closed (LD5), respectively, and both may promote the issue to be closed; 
(3) After the release of LD3 (Confirmation), the number of leadership increased, but the number of comments remained unchanged, indicating that it can promote the emergence of leadership, and it may increase the time for discussion;
(4) The discussions after LD4 (Inquiry) are more active; 
(5) LD6 (Volunteer) has no difference in most features, but in comparison, it is far from the opening and closing of the issue. We observed that it is generally in the late stage of the issue and will start the discussion of issue repair, so it takes longer to close the issue.

In summary, it can be concluded that LD3, LD4 play important role in stimulating issue discussion process, and LD1, LD2 contributes to bring speedy issue resolution.

\subsection{Human Evaluation}

To evaluate the usefulness of {\tool}, we recruited 9 developers from three OSS communities (i.e., \textit{bitcoin}, \textit{atom}, \textit{ember.js}) as evaluator.
Specifically: (1) use {\tool} to identify leadership behaviors in these three projects, and then derive a ranked list of developers according to the frequency of leadership behaviors;
(2) create two subsets of developers: one with the Top 20 developers (used as study group), and the other with developers ranking 21-40 (used as comparison group);
(3) randomly select 10 developers from each group, and ask the human evaluators (from the same OSS project) to vote (Yes or No) for the 20 randomly selected developers, based on whether they believe the names on the list play important role(s) in driving issue resolution progress. Note that the developers are  provided in random order (without ranking information) to the human evaluators to avoid introducing potential bias. 

The results show that the random developers from the first subset (i.e., Top-20) account for 74\% of overall votes, while those from the second subset receives 26\% of overall votes. This indicates that, human evaluators' votes are mostly consistent with the leadership status identified by {\tool}.

\section{Discussion}
\subsection{Practical Insights for OSS Communities}

Our findings suggest that leadership in OSS projects should encompass the key facets of emergent leadership behaviors in issue discussion. Based on the results from this study, we provide the following practical insights for OSS community. 

First, while developers with large code submissions are naturally considered as leaders/core contributors to the project, OSS projects may not necessarily pay attention to emergent leadership of other developers, who serve important influential roles in guiding/facilitating issue resolution. In addition to recognize project leaders based on their code contribution, OSS projects should also consider developers emergent leadership behaviors in their interpersonal communications, e.g. issue comments and live chat. 

Second, with support from {\tool}, OSS community can automate the following tasks accordingly: (1) to identify developers with the most influential emergent leadership behaviors according to the LD1 (Proposal) category; (2) to identify developers coordinating the most dominant brainstorming and exploration tasks according to the LD3 (Confirmation) and LD4 (Inquiry) categories; and (3) to identify developers facilitating the management and operation tasks such as redirecting a conversation, issue closing and volunteering, according to LD2 (Redirection), LD5 (Operation) and LD6 (Volunteer). 
With these knowledge, there is also great potential to appropriately automate and streamline such issue discussion and operational tasks, in order to accelerate issue resolution cycles.

\subsection{Practical Insights for OSS Developers}

We also observe several findings that may shed light to practices related to leadership skill development.
First, developers will benefit from learning about different roles of emergent leadership. To emerge and to be accepted as a leader in an OSS project, a developer needs to know how to communicate effectively with others via textual messages. As the core construct of {\tool}, the linguistic pattern set embodies strong verbal-cues correlated with the emergent leadership. These can be used to coach newcomer developers in effectively inquiring about an issue, collaboratively confirming a candidate solution, or clearly proposing an alternative option, etc.  

Second, while intuitively, it may seem that OSS leaders should play greater roles in LD1 (Proposal), since it is the most influential leadership behaviors, we actually observe something slightly different. We observe no much difference between junior and senior OSS developers in three types of leadership behaviors, i.e., LD1 (Proposal), LD4(Inquiry), and LD6(Volunteer). In addition, junior developers tend to conduct more LD3 (Confirmation), and less LD2 (Redirection) and LD5 (Operation); and senior developers conduct more LD2 (Redirection) and LD5 (Operation), and less LD3 (Confirmation). 
It is easy to understand that LD2 and LD5 rely on comprehensive knowledge about a specific OSS projects. 
Therefore, we encourage new or in-experienced developers to focus more on LD1, LD3, LD4, and LD6 when looking for opportunities to make contributions to a new OSS project.

\subsection{Threats to Validity}
\label{subsec:threats}

The first threat concerns the generality of the proposed approach. The {\tool} is only evaluated using 10 popular OSS projects, which might not be representative of other projects. 
However, the projects are from various domains, and the performance convergence experiment is conducted to ensure the comprehensiveness and effectiveness of the finalized linguistic rules.

The second threat comes from the representativeness of the six categories of emergent leadership behaviors. One co-author's area of research is in leadership, and the six leadership categories are carefully identified based on reviewing and examining both traditional leadership styles and emergent leadership studies for virtual teams. In addition, the leadership survey results also show that this provides a meaningful representation of emergent leadership behaviors in OSS issue discussion communications. 

The third threat is associated with the results of manual labelling. We follow an iterative process\cite{pustejovsky2012natural}, and conduct the labelling by three authors independently, and common consensus are reached after several rounds of discussion. 

The last threat is associated with the human evaluation since the number of participants is small. We cannot guarantee that the votes from the human evaluators are fair. To mitigate this threat, we ask the evaluators to vote on 10 randomly selected developers from two subsets, and use the total votes for comparison, to mitigate individual bias. 

\section{Related Work}
\label{sec:related_work}

\textbf{Linguistic Patterns.} Previous researches noticed that people tend to use recurrent expressions to illustrate similar things, and leverage such phenomenon, existing researches constructed the linguistic patterns for automating various software engineering tasks. Zhao et al.\cite{DBLP:conf/sigsoft/Zhao0BSWMW20} conducted a manual analysis of 980 performance issue reports and extracted 80 linguistic patterns; they then combined these patterns with machine/deep learning algorithms to automatically detect the performance issue reports. Lin et al.\cite{DBLP:conf/icse/0008ZBPL19} manually analyzed 388 API-related sentences and defined 157 linguistic patterns, and used these patterns to classify the opinions about API on Q\&A website. Shi et al.\cite{DBLP:conf/kbse/ShiCWLB17} proposed an linguistic patterns based approach to automatically understand the feature request. They manually defined a group of patterns and leverage them to generate fuzzy rules which can automatically identify feature request. Panichella et al.\cite{DBLP:conf/icsm/PanichellaSGVCG15} manually inspected 500 app reviews and defined 246 patterns to automatically classify the app reviews into categories relevant to software maintenance and evolution. Motivated by these studies, this paper utilized linguistic patterns to identify the leadership behaviors.

\textbf{Leadership in OSS Community.} The leadership behaviors have long been investigated in business and management domain, and have attracted increasing attention in software engineering domain in recent years. Giuri et al. \cite{Giuri2008explaining} characterized the difference in individual and project-specific characteristics between project leaders and other project members. Bian et al. \cite{bian2018online} explored the correlation between leader characteristics and success of OSS projects from the behavioral, structural and cognitive dimensions. Li et al. \cite{Li2012leadership} investigated the relationship between leadership characteristics and the motivation to contribute. Jensen et al. \cite{jensen2005collaboration} explored the collaborative efforts, the leadership and control structures in OSS community. Tan et al. and Zhang et al. \cite{tan2019how,tan2020scaling,zhang2020how} investigated how developers and companies collaborate and communicate with each other during OSS development. By comparison, this paper proposes an automatic approach to identify leadership behaviors in OSS projects, with which one can conduct the leadership related investigation to facilitate the OSS development.

\section{Conclusion}
\label{sec:conclusion}

To facilitate the developing and sustaining health growth of the OSS communities, 
this paper explores an innovative perspective to look at OSS leadership, and proposes an automated approach, {\tool}, to mine communication styles and identify emergent leadership behaviors in OSS projects using issue comments data. 
Based on the constructed six categories of leadership and extracted heuristic linguistic patterns, {\tool} employs an automated algorithm to consolidate the pattern ranking list for leadership identification.
The evaluation results demonstrate the effectiveness of the lightweight heuristic rule-based approach employed in {\tool}, outperforming 10 Machine Learning baseline models. Applying the proposed {\tool} to 5 OSS projects, we conduct comparison of identified leadership with existing leadership indicators and offer practical insights on community building and leadership skill development. This study provides a new lens for examining leadership and its influence in OSS projects. With the support of {\tool, one can} look at how leadership develops, at who tends to become a leader, how leadership evolves in a developer’s career, etc.

Future researches include: 1) further evaluate and improve the performance of {\tool} using more OSS projects; 2) investigate the impact of various emergent leadership behaviors; 3) examine the relationship between emergent leadership and temporal, contextual structure of OSS communities;
4) develop intelligent support for facilitating leadership skill development of OSS developers.


\bibliographystyle{ACM-Reference-Format}
\bibliography{reference}


\end{document}